\documentclass[amsmath,amssymb,preprintnumbers,nofootinbib,prd,a4paper,twocolumn]{revtex4-1}
\pdfoutput=1
\usepackage{amsthm}
\usepackage{amsmath}
\usepackage{graphicx}
\usepackage{bbold}
\usepackage{color}
\usepackage{subfigure}
\usepackage{multirow}
\usepackage{graphicx,bm}
\usepackage{physics}
\usepackage[vcentermath]{youngtab}

\usepackage{dcolumn}
\usepackage{bm}
\usepackage{slashed}
\usepackage{epsfig}
\usepackage[colorlinks=true, linkcolor=black, citecolor=black, urlcolor=black]{hyperref}
\usepackage{verbatim}
\usepackage[bordercolor=gray!20,backgroundcolor=blue!10,textsize=tiny]{todonotes}

\usepackage[utf8]{inputenc}
\usepackage[T1]{fontenc}
\usepackage[normalem]{ulem}

\newcommand{\ii}{\mathrm{i}}
\newcommand{\beq}{\begin{eqnarray}}
\newcommand{\eeq}{\end{eqnarray}}

\newcommand{\bmp}{\noindent\begin{minipage}{16cm}}
\newcommand{\emp}{\end{minipage}\vskip 7mm} 

\newcommand{\be}{\begin{eqnarray}}
\newcommand{\ee}{\end{eqnarray}}

\newcommand{\SU}{\mbox{SU}}
\newcommand{\SO}{\mbox{SO}}
\newcommand{\SP}{\mbox{Sp}}
\newcommand{\UU}{\mbox{U}}

\newcommand{\st}{s_\theta}
\newcommand{\ct}{c_\theta}

    \newcommand{\fL}{f_{\Lambda}}
    
    \newcommand{\cQ}{c_{Q}}
    \newcommand{\cL}{c_{\Lambda}}

\begin{document}
\title{Composite self-interacting dark matter and Higgs}

\author{Martin {\sc Rosenlyst}}
\email{martin.jorgensen@physics.ox.ac.uk}
\affiliation{Rudolf Peierls Centre for Theoretical Physics, University of Oxford, 1 Keble Road, Oxford
OX1 3NP, United Kingdom}


\begin{abstract}
	
We propose a novel mechanism in composite models that provides self-interacting dark matter along with the Higgs itself as composite particles, alleviating the Standard Model naturalness problem and explaining small-scale discrepancies such as the core-cusp and ``too big to fail'' problems. These dark matter candidates are stable due to global $ \UU(1) $ symmetries of the composite dynamics and their strong self-interactions are created by the novel mechanism based on top-quark partial compositeness. The relic density of the dark matter candidates is particle anti-particle symmetric and due to thermal freeze-out. We implement this mechanism in a four-dimensional gauge theory with a minimal number of fermions charged under a new confining gauge group $G_{\rm HC} $.

\end{abstract}

\maketitle


The overwhelming evidence from astronomical and cosmological observations~\cite{Planck:2018vyg} suggests dark matter (DM) as the dominant form of matter in the Universe making up about $ 85\% $ of its mass. Understanding the non-baryonic nature of DM is arguably one of the most important mysteries in our knowledge of the physical world. The last decades, the main DM paradigm, Weakly Interacting Massive Particles (WIMPs)~\cite{Steigman:1984ac}, is currently challenged by the non-observation of the many WIMP candidates, though advances in collider, direct-detection and indirect-detection experiments. Similarly, many DM density profiles near galactic centers and in cluster halos, inferred from observations, appear cored (shallow) rather than the cuspy (steep) profiles expected from N-body simulations of halo formation with WIMPs~\cite{Navarro:1996bv,Navarro:1996gj,Borriello:2000rv,deBlok:2002vgq,954985,2010AdAst2010E}. This is referred to as the core-cusp problem~\cite{1994Nature,1994ApJ,2008ApJ,2011ApJ74220W,2012MNRAS419184A,2012ApJ754L39A,2014ApJ78963A,Oh:2015xoa}, indicating that large self-interactions among DM particles themselves, which is known to produce such cores~\cite{Spergel:1999mh,Kamada:2016euw}, could be the dominant DM particle interactions. Furthermore, simulations with only collisionless DM predict a substantial population of massive concentrated subhaloes, which is incompatible with the stellar kinematics of observed satellite galaxies around, e.g., the Milky Way and Andromeda Galaxy~\cite{2011MNRAS,2012MNRAS4221203B,Tollerud:2014zha}, known as the ``too big to fail" (TBTF) problem. This problem also arises in dwarf galaxies in Local Group~\cite{2014MNRAS444222G,Kirby:2014sya} and beyond~\cite{Papastergis:2014aba}. 

The work in Refs.~\cite{Spergel:1999mh,Dave:2000ar} and in the review of Ref.~\cite{Weinberg:2013aya} show that the core-cusp and TBTF issues can be ameliorated by Self-Interacting DM Particles (SIDMs)~\cite{1992ApJ,Hochberg:2014dra} with strong elastic contact-type scattering among themselves, with velocity-independent cross-section per mass in the range $ \sigma_{SI}/m_{DM}\approx 0.5-1~\text{cm}^2/\text{g} $~\cite{Tulin:2017ara}. However, according to studies of merging galaxy clusters observations~\cite{Clowe:2006eq,Randall:2008ppe,2012ApJ747L42D}, halo sphericity~\cite{Peter:2012jh} 
and massive clusters~\cite{Kaplinghat:2015aga,2018ApJ853109E,Sagunski:2020spe}, weaker self-interactions are preferred for these systems, with cross-section $ \sigma_{SI}/m_{DM}\sim \mathcal{O}(0.1)~\text{cm}^2/\text{g} $. Thus, these studies favor velocity-dependent cross-section that decreases with increasing scattering velocity, because the typical relative velocities between DM particles are higher in more massive haloes. 

However, recent studies of ultra-faint dwarf galaxies (UFDs)~\cite{Hayashi:2020syu} provides constraints on the self-interacting cross-section of $ \sigma_{SI}/m_{DM} \lesssim \mathcal{O}(0.1)~\text{cm}^2/\text{g} $. This discrepancy is difficult to explain with velocity-dependent cross-section, as the typical velocities of the DM in the UFDs and in the dwarf irregular galaxies are close with each other. Therefore, the larger cross-section preferred by the dwarf irregular and the low surface brightness galaxies may be explained by baryonic feedback effects~\cite{Navarro:1996bv,Gnedin:2001ec,Read:2004xc,Governato:2009bg,2011ApJ741L29K}, since these systems are more affected by those effects than the UFDs. The different mechanisms of SIDM and baryonic physics in galaxies may coincide and interplay such that all the discrepancies can be alleviated by introducing a velocity-independent, elastic scattering cross-section. Finally, according to SIDM-baryon simulations of, e.g., dwarf galaxies~\cite{Vogelsberger:2014pda}, Milky-Way-mass galaxies~\cite{Vargya:2021qza} and galaxy clusters~\cite{Robertson:2017mgj}, it may be possible in the future to set limits on the SIDM cross-section for these systems by a better resolution of the mass distribution and characteristics of their central baryonic matter. 

In the Standard Model (SM), we know of large self-interactions among the composite hadrons via the strong nuclear force. It is, therefore, natural to imagine and investigate the hypothesis that DM-DM self-interactions
arise from a new, but analogous type of composite dynamics. The search for composite
DM is motivated by the fact that the bulk of visible mass in the universe arises from composite dynamics. Even more remarkably, this hypothesis is further motivated by that the SM description of the origin of visible mass, the elementary Higgs particle, is plagued by naturalness and triviality problems, which may be alleviated if the Higgs particle itself is composite~\cite{Terazawa:1976xx,Terazawa:1979pj,Kaplan:1983fs,Kaplan:1983sm}. We are, therefore, lead to investigate a possible common composite dynamics origin of the Higgs and SIDM
-- for so-called Composite Higgs (CH) models~\cite{Kaplan:1983fs} -- for which the CH research so far has most dominantly focused on the Higgs mechanism and composite DM candidates with weak or no self-interactions.

This letter presents a first proposal of composite models that predicts DM with strong self-interactions along with the Higgs itself as composite pseudo-Nambu-Goldstone bosons (pNGBs) from a common composite dynamics, allowing for a successful description of the electroweak (EW) symmetry and SIDM.~\footnote{This is realized in the largely excluded Technicolor model~\cite{Frandsen:2011kt} with the Higgs boson as a heavy composite scalar excitation.} Our proposal features the following key ingredients:  \begin{itemize}
	\item[i)] a SM-like composite pNGB Higgs multiplet with custodial symmetry~\cite{Georgi:1984af}; 
	\item[ii)] a composite DM candidate, stable due to a global symmetry of the new strong interactions;
	\item[iii)] an annihilation mode of the DM, resulting in a thermal relic density;
	\item[iv)] strong elastic contact-type scattering
among DM particles themselves, with cross-section per mass
$ \sigma_{SI}/m_{DM}\sim 0.1-1~\text{cm}^2/\text{g} $.
\end{itemize} The first ingredient i) is the main feature of CH models~\cite{Kaplan:1983fs}, Little Higgs~\cite{ArkaniHamed:2001nc,ArkaniHamed:2002qx}, holographic extra dimensions~\cite{Contino:2003ve,Hosotani:2005nz}, Twin Higgs~\cite{Chacko:2005pe} and elementary Goldstone Higgs models~\cite{Alanne:2014kea}. In the CH framework, the second ingredient ii) has been accommodated by extending the global symmetry, typically producing thermal DM candidates~\cite{Ma:2015gra,Wu:2017iji,Cai:2018tet,Alanne:2018xli,Cacciapaglia:2019ixa} satisfying i)-iii) and in few model types non-thermal candidates~\cite{Cai:2019cow,Cai:2020bhd,Cacciapaglia:2021aex}. However, until now, none of these model types satisfies iv).

To realize this, we propose for concreteness four-dimensional gauge theories with a single strongly interacting gauge group $G_{\rm HC}$, with $N_1$ Weyl fermions (denoted $ Q_1 $) in the representation $\mathcal{R}_1$ and $N_2$ Weyl  fermions ($ Q_2 $) in the representation $\mathcal{R}_2$. The $\mathcal{R}_1$ fermions are gauged under the EW interactions, while the $R_2$ fermions are inert under the SM gauge symmetry for simplicity.~\footnote{The $R_2$ fermions may be charged under some new gauge symmetries, for example, an $ \SU(2)_{\rm R} $ gauge symmetry as in Ref.~\cite{Cacciapaglia:2021aex}.} Upon the compositeness scale, the $ \mathcal{R}_1 $ fermions form the composite Higgs doublet with the decay constant $ f_{Q_1} $ accommodating i), while the $ \mathcal{R}_2 $ fermions with $ f_{Q_2} $ result in composite DM candidates stable due to $ \UU(1) $ symmetries, satisfying ii). However, it may be possible to obtain a DM candidate consisting of $ \mathcal{R}_2 $ fermions charged under the SM gauge symmetry, but for such a model, it turns out to be challenging to implement the novel mechanism that provides the strong self-interactions of DM. Furthermore, there exists a global $\UU(1)_\Theta$ symmetry under which both sets of hyper-fermions are charged, which is spontaneously broken by the condensates and generates a singlet pNGB, $ \Theta $, with the decay constant $ f_\Theta $. The kinetic term of $ \Theta $ is canonically normalized if $ f_{Q_1}=f_{Q_2}=f_\Theta\equiv f $, which we will assume for simplicity. Generally, the kinetic terms must be renormalized, but we expect them to be of the same size based on Casimir scaling~\cite{Ryttov:2008xe,Frandsen:2011kt,Alanne:2018xli}. Finally, we add a new class of interactions that both dynamically induce SM-fermion masses, align the vacuum of the $ \mathcal{R}_1 $ fermions into the CH regime, and generate the annihilation mode and strong self-interactions of the DM candidates. We will discuss fermion partial compositeness (PC) type~\cite{Kaplan:1991dc} as an example of such interactions, providing the ingredients iii) and iv). In that case, we add a third sector with QCD charged fermions to accommodate top partners. 

To assure condition iv) and generate strong enough self-interactions, the DM candidates require masses of $ m_{DM}\lesssim \mathcal{O}(1)~\text{GeV}$, which are much smaller than the Higgs decay constant of $ f \gtrsim \mathcal{O}(1)~\text{TeV}$. This lightness of the DM candidates entails no fulfillment of condition iii) due to kinematics since the remaining composite states, with typical masses of about $\mathcal{O}(f) $ (except for the Higgs boson), are much heavier than these DM candidates. However, in the models considered, there typically exists a pNGB singlet, $ \eta $, in the sector of the $ \mathcal{R}_1 $ hyper-fermions that can mix with the pNGB singlet $ \Theta $ by supposing that only one $ \mathcal{R}_1 $ hyper-fermion doublet achieves an explicit vector-like mass. In that case, the two pNGBs $ \Theta $ and $ \eta $ will almost mix half-and-half, leading to a very light singlet consisting mostly of $ \Theta $ along the lines of Refs.~\cite{Ferretti:2016upr,Belyaev:2016ftv}. It may allow the DM candidates to annihilate into these light singlets and thermally produce the relic density. We, therefore, need an efficient annihilation mode of the DM candidates into the light $ \Theta $. 

This can be realized by adding PC four-fermion operators of the form $ q_{L,3} Q_1 Q_1 \chi $ and $ t_R Q_2 Q_2 \chi $, where $ q_{L,3} $ and $ t_R $ are, respectively, the third generation of the left-handed quark doublet and the right-handed top singlet. Further, $\chi$ is a new hyper-fermion, transforming in a suitable representation of $ G_{\rm HC}$, and carrying appropriate quantum numbers under the SM gauge symmetry. In traditional CH models, the left- and right-handed top partners only consist of one hyper-fermion type $ Q $ in one representation of $ G_{\rm HC}$ together with $ \chi $, where the PC operators are of the form $ q_{L,3} Q Q \chi $ and $ t_R Q Q \chi $. Besides that these new non-traditional PC operators dynamically induce the top mass and top-Yukawa coupling, the two sectors of the $ \mathcal{R}_{1,2} $ fermions will be linked together due to the mixing of the two top partners, $ Q_1Q_1 \chi $ and $ Q_2Q_2 \chi $. This results in an annihilation mode of the DM candidates, contained in the $ Q_2 $--sector, into $ \eta $ states, in the $ Q_1 $--sector. Due to the nearly half-and-half mixing of $ \eta $ and $ \Theta $, there exists an annihilation channel of the DM candidates into light $ \Theta $ states, fulfilling condition iii). Furthermore, via the mixing of these non-traditional right-handed top partners, these PC operators also provide strong self-interactions of the DM candidates with quartic couplings of order unity, assuring condition iv). For traditional PC operators, those self-interactions are unfortunately suppressed by $ (m_{DM}/f)^2 $ with $ m_{DM}\ll f $.   

\begin{table}[t]
    \begin{center}
	\begin{tabular}{ccccc}
	    \hline
	    & $\vphantom{\frac{\frac12}{\frac12}}\quad G_{\mathrm{HC}}\quad$ & $\quad\SU(2)_{\mathrm{W}}\quad$ 
    & $\quad\mathrm{U}(1)_{\mathrm{Y}}\quad$ & $\quad\mathrm{U}(1)_{\Lambda}\quad$ \\
	    \hline
	    $\vphantom{\frac{\frac12}{\frac12}} (U_L,D_L)$	&   ${\tiny \yng(1)}$	&   ${\tiny \yng(1)}$	&   0 & 0\\    
	    $\vphantom{\frac{1}{\frac12}} \widetilde{U}_{\mathrm{L}}$	&   ${\tiny \yng(1)}$	&   1	&   $-1/2$ & 0\\    
	    $\vphantom{\frac{1}{\frac12}} \widetilde{D}_{\mathrm{L}}$	&   ${\tiny \yng(1)}$	&   1	&   $+1/2$  & 0
	    \\  
	   $\vphantom{\frac{1}{\frac12}} \lambda_{ \mathrm{L}}$	&   ${\rm Adj}$	&   1	&   $0$ & $ +1/2 $ \\    
	   $\vphantom{\frac{1}{\frac12}} \widetilde{\lambda}_{ \mathrm{L}}$	&   ${\rm Adj}$	&   1	&   $0$ & $ -1/2 $ \\  
	    \hline
	\end{tabular}
    \end{center}
        \caption{The new fermion content and their charges, as in Refs.~\cite{Ryttov:2008xe,Alanne:2018xli}. All groups are gauged except for
$ \UU(1)_\Lambda $, which is a global symmetry in the $ \Lambda $ sector responsible for dark matter stability.}
    \label{table:fullmodel}
\end{table}

For concreteness, for one of the hyper-fermion sectors with a single representation, the symmetry breaking patterns are known~\cite{Witten:1983tx,Kosower:1984aw}: Given $N$ Weyl fermions transforming as the $\mathcal{R}$ representation of $ G_{\rm HC} $, the three possible classes of vacuum cosets are $ \SU(N)/\SO(N) $ for real $\mathcal{R}$, $ \SU(N)/\SP(N) $ for pseudo-real $\mathcal{R}$ and $ \SU(N)\otimes \SU(N)\otimes \UU(1)/ \SU(N)\otimes \UU(1)$ for complex $R$~\cite{Peskin:1980gc}. For these three classes, the minimal CH cosets that fulfill requirement i) contain $N=5$ in the real case~\cite{Dugan:1984hq}, and $N=4$ in both the pseudo-real~\cite{Galloway:2010bp} and the complex cases~\cite{Ma:2015gra}. In terms of pNGB spectrum, the pseudo-real case is the most minimal, with only five states. Similarly, the minimal DM cosets that can fulfill the condition ii) contain $N=2$ in the real~\cite{Ryttov:2008xe}, $N=4$ in the pseudo-real~\cite{Ryttov:2008xe} and $N=2$ in the complex case~\cite{Kaplan:1983fs}. We will, therefore, focus on the real case since it only needs two Weyl fermions and has the most minimal pNGB coset, with one complex pNGB singlet playing the role of DM candidate. In the following, these two minimal subsets will be combined in one minimal model example, fulfilling the requirements i) and ii), while iii) and iv) are assured by implementing non-traditional PC operators. 

From now on, we focus on the minimal model example that fulfills all the conditions i)-iv). In this minimal model example, the gauge group can be chosen to be the minimal $G_{\mathrm{HC}}=\SU(2)_{\mathrm{HC}}$ with $\mathcal{R}_1$ as the fundamental representation and $N_1=4$ Weyl spinors, arranged in one $\SU(2)_{\rm L}$ doublet $(U_L,D_L)$ and two singlets $\widetilde{U}_L$ and $ \widetilde{D}_L$. We further take $\mathcal{R}_2$ to be the adjoint representation and $N_2=2$ Weyl spinors, arranged in two SM singlets $\lambda_L$ and $ \widetilde{\lambda}_L $, as in Refs.~\cite{Ryttov:2008xe,Alanne:2018xli}. The fermion content in terms of left-handed Weyl fields, with
$ \widetilde{\psi}_{\mathrm{L}}\equiv \epsilon \psi_R^* $, along with their EW quantum numbers are presented in Table~\ref{table:fullmodel}. We provide only vector-like mass to either the hyper-fermion pair $ U_L,D_L $ or $ \widetilde{U}_L,\widetilde{D}_L $ because it helps stabilize the CH vacuum~\cite{Cacciapaglia:2014uja,Dong:2020eqy} while keeping the pNGB $ \Theta $ light (see the Supplementary Material for detail). Finally, the spinors $Q=(U_{\mathrm{L}},\, D_{\mathrm{L}},\, \widetilde{U}_{\mathrm{L}},\, \widetilde{D}_{\mathrm{L}})$ 
and $\Lambda=(\lambda_{ \mathrm{L}}, \,   \widetilde{\lambda}_{ \mathrm{L}})$
transform in the fundamental representations of the $\SU(4)_Q$ and $\SU(2)_\Lambda$ subgroups of the global symmetry, respectively, where we drop an L subscript on $Q$ and $\Lambda$ for simplicity. 

Below the condensation scale, $ \Lambda_{\rm HC}\sim 4\pi f $, this minimal model can be described in terms of an effective theory, following the CCWZ prescription~\cite{Marzocca:2012zn}, and based uniquely on the coset symmetry: 
\begin{equation} \label{eq: coset symmetry}
\frac{\SU(4)_{Q} \otimes \SU(2)_{\Lambda} \otimes \UU(1)_\Theta}{\SP(4)_{Q} \otimes \UU(1)_{\Lambda}\otimes \mathbb{Z}_2}\,.
\end{equation} The minimal $ \SU(4)_Q/\SP(4)_Q $ coset contains the longitudinal components of the $W^\pm$ and $Z$ bosons, a Higgs candidate $ h $ and a singlet $\eta$, while the $ \SU(2)_\Lambda/\SO(2)_\Lambda\cong \SU(2)_\Lambda/\UU(1)_\Lambda $ coset includes the DM candidate as a complex pNGB state $ \Phi $. In addition, the above coset contains the pNGB singlet $ \Theta $ corresponding to the spontaneous breaking of the $ \UU(1)_\Theta $ symmetry. The DM candidate $ \Phi $ is the only pNGB state charged under the unbroken global $ \UU(1)_\Lambda $ symmetry, while all the pNGB states are even under the unbroken $ \mathbb{Z}_2 $ symmetry. 

The remaining essential ingredient for model building is the operators that provide the SM-fermion masses and help align the vacuum of the $ Q $--sector. As already mentioned, non-traditional operators are also crucial for generating an annihilation mode of and self-interactions among the DM candidate, $ \Phi $, in the $ \Lambda $--sector. These operators can be written as PC-type~\cite{Kaplan:1991dc} as follows \begin{equation} \begin{aligned}\label{eq: PC operator top yukawa}
\frac{\widetilde{y}_L}{\Lambda_{t}^2}q_{{\rm L},3}^\alpha (Q^T P_{Q}^\alpha Q \chi)+ \frac{\widetilde{y}_R}{\Lambda_{t}^2}t_{\rm R}^c (\Lambda^T P_{\Lambda} \Lambda \chi )+{\rm h.c.}\,,
\end{aligned} \end{equation} where \begin{equation} \begin{aligned}
&(P_Q^1)_{ij}=\frac{1}{\sqrt{2}}(\delta_{i1}\delta_{j3}-\delta_{i3}\delta_{j1})\,,\nonumber \\ &(P_Q^2)_{ij}=\frac{1}{\sqrt{2}}(\delta_{i2}\delta_{j3}-\delta_{i3}\delta_{j2})\,,\nonumber \\ &(P_\Lambda)_{ij}=\frac{1}{\sqrt{2}}(\delta_{i1}\delta_{j2}+\delta_{i2}\delta_{j1})\nonumber
\end{aligned} \end{equation} for the spurions for the left-handed top ($ P_Q^\alpha $) transforming as the two-index anti-symmetric representation of the chiral symmetry $ \SU(4)_Q $ and the spurion for the right-handed top ($ P_\Lambda $) transforming as the two-index symmetric representation of $ \SU(2)_\Lambda $. For these operators, the underlying theory has to be extended with extra QCD charged fermions, $ \chi $, in order to construct the above top partners. Furthermore, the minimal gauge group $ G_{\rm HC}=\SU(2)_{\rm HC} $ needs to be enlarged to $ \SP(2N)_{\rm HC} $, ensuring asymptotic freedom as studied in, for example, Ref.~\cite{Ferretti:2013kya}. 

One possible extension of our model can be to introduce six new $ \chi $ Weyl fermions, transforming under the fundamental representation of QCD and two-index anti-symmetric of $ G_{\rm HC} $, as detailed in Ref.~\cite{Ferretti:2014qta}. Moreover, for this model example, the four $ Q $ and two $ \Lambda $ Weyl fermions are still in the fundamental and adjoint representation of $ G_{\rm HC} $, respectively. Asymptotic freedom is guaranteed by $ N<3 $, so we choose $ G_{\rm HC}=\SP(4)_{\rm HC} $. In this scenario, the global symmetry in Eq.~(\ref{eq: coset symmetry}) is extended by the coset $ \SU(6)_\chi/\SO(6)_\chi $ and one extra $ \UU(1)_\Theta $ symmetry. In our case, the pNGB associated with the spontaneous breaking of the extra $ \UU(1)_\Theta $ symmetry does not mix with neither $ \eta $ nor $ \Theta $. Therefore, it does not influence the lightness of the $ \Theta $ state (see the supplementary material). Finally, we must investigate whether it is possible to contract the HC indices in Eq.~(\ref{eq: PC operator top yukawa}), leading to HC invariant top partners. The fundamental (for $ Q $), two-index anti-symmetric ($ \chi $) and adjoint ($ \Lambda $) representations of $ \SP(4)_{\rm HC} $ are $ \mathbf{4} $, $ \mathbf{5} $ and $ \mathbf{10} $, respectively. Since $ \mathbf{4}\otimes \mathbf{4}\otimes \mathbf{5} = \mathbf{1}\oplus\dots $ and $ \mathbf{10}\otimes \mathbf{10}\otimes  \mathbf{5} = \mathbf{1}\oplus \dots $ according to Ref.~\cite{Feger:2019tvk}, the left- and right-handed top partners of the form $ Q^T P_Q^\alpha Q \chi $ and $ \Lambda^T P_\Lambda \Lambda \chi $, respectively, can be contracted such that they are invariant under $\SP(4)_{\rm HC} $. 

Upon the compositeness scale, the effective Lagrangian has the form
\begin{equation} \label{eq:Left}
\mathcal{L}_{\rm EFT} = \mathcal{L}_{\chi pt} - V_{\rm eff}\,,
\end{equation}
where the first term corresponds to the usual chiral perturbation theory for the pNGBs, and the second term contains the effective potential generated by loops of the EW gauge bosons and the fermions (dominated by the top loops). As already mentioned, the effective potential plays a crucial role in determining the physics of the composite SIDM candidate and the EW symmetry breaking via the vacuum alignment (see the Supplementary Material for detail).

So far, the conditions i) and ii) are fulfilled by the above model example by providing a SM-like Higgs and a DM candidate stabilized by the unbroken $ \UU(1)_\Lambda $ symmetry. To satisfy condition iii), the DM pNGB $ \Phi $ needs to annihilate appropriately to provide the correct relic density from thermal freeze-out. Because the DM candidate is light ($ \lesssim \mathcal{O}(1) $~GeV) relative to the decay constant ($ \gtrsim \mathcal{O}(1) $~TeV), the only annihilation channel is in a pair of light $ \Theta $ pNGB states, which has a cross-section of the form: \begin{equation} \begin{aligned}\label{Eq: sigma v DM DM to eta eta}
&\langle  \sigma v\rangle_{\Theta} = \frac{\lambda_{\Phi\Phi\Theta\Theta}^2}{32\pi  m_{DM}^2}\sqrt{1-\frac{m_{\Theta}^2}{m_{DM}^2}}\,.
\end{aligned}\end{equation}
The quartic coupling $ \lambda_{\Phi\Phi\Theta\Theta} $ is generated via the mixing of the top partners in Eq.~(\ref{eq: PC operator top yukawa}), resulting in a connection between the cosets containing the DM candidate and the pNGB singlet $ \eta $, respectively; meanwhile, $ \eta $ is almost half-and-half mixed with $ \Theta $. The misalignment in the $ Q $--coset suppresses this coupling, $\lambda_{\Phi\Phi\Theta\Theta} \approx \lambda_0 (v_{\rm SM}/f)^2$. The explicit expression of $ \lambda_0 $ for the minimal model example is shown in the Supplementary Material, which is optimal order of unity. This process can provide the thermal DM relic density $ \langle  \sigma v\rangle_{\Theta} \simeq 5\times 10^{-26}~\text{cm}^3~\text{s}^{-1}$ for $ m_{DM}\lesssim 1~\text{GeV} $~\cite{Steigman:2012nb}, implying 
\begin{equation} \label{DMmass cond}
m_{DM} \simeq 0.9~\text{GeV} \times \left(\frac{10~\text{TeV}}{f}\right)^2 \lambda_0
\end{equation}
for $m_\Theta \ll m_{DM}$ in Eq.~(\ref{Eq: sigma v DM DM to eta eta}).

\begin{figure}[t!]
	\centering
	\includegraphics[width=0.39\textwidth]{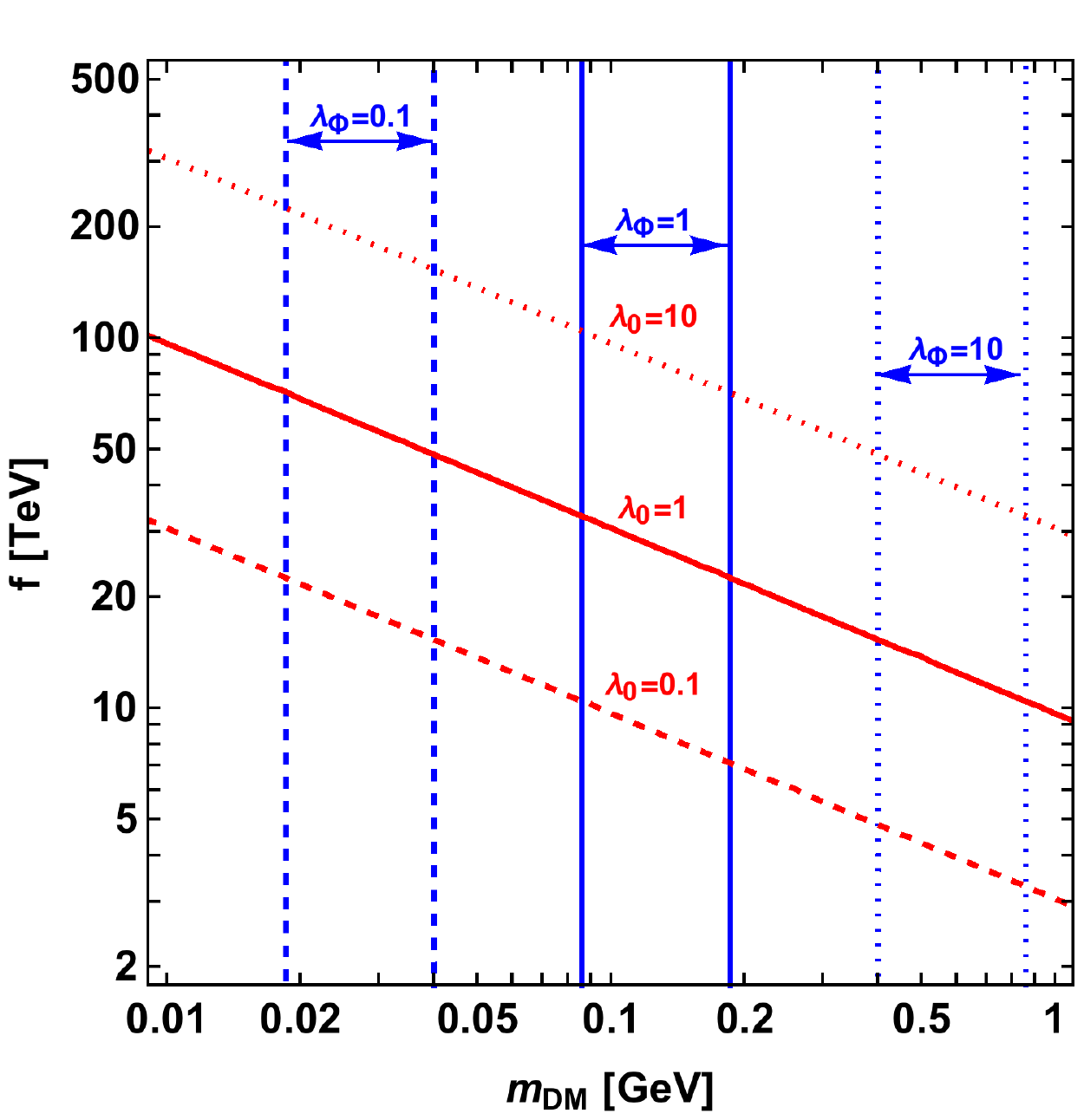} 
	\caption{Constraints on $ m_{DM} $ and the compositeness scale $ f $. On the red lines, we have a correct thermal relic density. These lines depends on $ \lambda_0 $, and we show three sample values: $ \lambda_0=0.1,1,10 $. Furthermore, an upper and lower limit on $ m_{DM} $ shown by the blue lines come from the constraints on the DM self-interacting cross-section per mass in Ref.~\cite{Tulin:2017ara}: $ \sigma_{ SI}/m_{DM}\gtrsim 0.1 $ and $ \sigma_{ SI}/m_{DM}\lesssim 1 $, respectively. These limits depend on $ \lambda_\Phi $ also illustrated by three sample values: $ \lambda_\Phi=0.1,1,10 $.
}
	\label{fig: Model Constraints}
\end{figure}

Finally, to ensure condition iv), the DM candidate must strongly self-interact with itself to alleviate small-scale discrepancies such as the core-cusp and TBTF problems, with a cross-section per mass in the range $ \sigma_{SI}/m_{DM}\sim 0.1-1~\text{cm}^2/\text{g}$~\cite{Tulin:2017ara}. Introducing the right-handed top partners in Eq.~(\ref{eq: PC operator top yukawa}), the mixing of these top partners generates elastic scattering of DM between themselves with a quartic coupling $ \lambda_{\Phi} $ in the order of unity  and thus a cross-section per mass of the form: \begin{equation} \begin{aligned}\label{Eq: SI cross section}
&\frac{\sigma_{SI}}{m_{DM}}=\frac{9 \lambda_{\Phi}^2}{\pi}\left(\frac{\text{GeV}}{m_{DM}}\right)^3\cdot 2\times 10^{-4}~\text{cm}^2/\text{g}\,.
\end{aligned}\end{equation}

For concreteness, we consider $ \lambda_0=\lambda_{\Phi}=1 $ in the following. According to the expressions in Eq.~(\ref{DMmass cond}) and~(\ref{Eq: SI cross section}), the conditions iii) and iv) are satisfied by a DM mass and Higgs decay constant in the intervals: \begin{equation} \begin{aligned}\label{Eq: benchmark point}
&90~\text{MeV}\lesssim m_{DM}\lesssim 190~\text{MeV}\,, \\ 
&\phantom{...}22~\text{TeV} \lesssim f\lesssim  32~\text{TeV}\,.
\end{aligned}\end{equation} Thus, the requirements i)-iv) can all be fulfilled by this novel mechanism based on partial compositeness. This low DM mass can be achieved by properly tuning some couplings in the effective potential in Eq.~\eqref{eq:Left}. This lightness is technically natural according to 't Hooft naturalness principle~\cite{tHooft:1979rat}, since a global symmetry is restored in the limit of a massless DM GB. Moreover, a value of $ f $ in the above interval leads to the pNGB Higgs potential requiring fine-tuning at the level of one part in $ (8-17) \cdot 10^3$.~\footnote{The tuning of the vacuum misalignment may be technically natural explained by the mechanism proposed in Ref.~\cite{Rosenlyst:2021tdr}.} However, this should not discourage the study of this model, as it represents a huge improvement compared to the fine-tuning in the SM, where the Higgs mass is fine-tuned to one part in $10^{34}$ against the Planck scale. \\

In summary, we have presented the first proposal of Composite Higgs models that both provide a composite pNGB Higgs and self-interacting DM from a common composite dynamics. In these models, a novel mechanism involving non-traditional top partners dynamically generates the SM-fermion masses, the correct vacuum misalignment, a thermal relic density of and strong self-interactions among the DM candidates. Fig.~\ref{fig: Model Constraints} shows an illustrative example of the allowed parameter space. This shows model-independently the main features of these models: For the couplings in the range $ 0.1 \leq \lambda_0,\lambda_\Phi \leq 10 $, a successful thermal relic density and self-interacting cross-section per mass can be obtained for DM masses in the interval $20~\text{MeV} \lesssim m_{DM}\lesssim 1000~\text{MeV} $ and for the Higgs decay constant $ 3~\text{TeV}\lesssim f \lesssim 200~\text{TeV} $. This novel mechanism of the Composite Higgs framework can thereby alleviate the SM naturalness, core-cusp, TBTF and maybe the remaining small-scale problems with a minimum of tuning, although some tuning remains necessary to obtain the necessary vacuum alignment.

\section*{Acknowledgements}
I acknowledge the funding from The Independent Research Fund Denmark, grant number DFF 1056-00027B. I would like to thank the Rudolf Peierls Centre for Theoretical Physics at the University of Oxford for hosting me during my two-year DFF-International Postdoctoral project. Finally, I would like to thank Mads~T.~Frandsen for useful discussions and comments on
the manuscript. 

\bibliographystyle{JHEP-2-2}

\bibliography{RTC-LCH.bib}

\clearpage
\onecolumngrid

\renewcommand{\thesubsection}{S\arabic{subsection}}
\setcounter{section}{0}
\renewcommand{\thefigure}{F\arabic{figure}}
\setcounter{figure}{0}
\renewcommand{\thetable}{T\arabic{table}}
\setcounter{table}{0}
\renewcommand{\theequation}{E\arabic{equation}}
\setcounter{equation}{0}

\begin{center}
	\Large 
\textbf{Supplementary material}
\end{center}

\section{The electroweak embedding and The condensates}

Here we consider in more detail the model example with $ \mathcal{R}_1 $ pseudo-real and $ \mathcal{R}_2 $ real of $ G_{\rm HC} $ studied in the main text. To describe the general vacuum alignment in the effective Lagrangian, we identify an $\SU(2)_{\rm L}\otimes\SU(2)_{\rm R}$ subgroup in $\SU(4)_Q$ by the left and right generators
\begin{equation}
    \label{eq:gensLR}
    T^i_{\mathrm{L}}=\frac{1}{2}\left(\begin{array}{cc}\sigma_i & 0 \\ 0 & 0\end{array}\right)\,, \quad \quad
    T^i_{\mathrm{R}}=\frac{1}{2}\left(\begin{array}{cc} 0 & 0 \\ 0 & -\sigma_i^{T}\end{array}\right)\,,
\end{equation}
where $\sigma_i$ are the Pauli matrices.  The EW subgroup is gauged after identifying the generator of 
hypercharge with $T_{\mathrm{R}}^3$. The alignment between the EW subgroup and the stability group $\SP(4)_Q $ can then be conveniently parameterized by an angle, $\theta$, after identifying the vacua that leave the 
EW symmetry intact, $E_Q^{\pm}$, and the one breaking it completely to $\mathrm{U}(1)_{\mathrm{EM}}$ 
of electromagnetism, $E_Q^{\mathrm{B}}$, with:
\begin{equation}
    E_{Q}^{\pm} = \left( \begin{array}{cc}
	\ii \sigma_2 & 0 \\
	0 & \pm \ii \sigma_2
    \end{array} \right)\,,\quad\quad
    E_{Q}^B  =\left( \begin{array}{cc}
	0 & \mathbb{1}_2 \\
	-\mathbb{1}_2 & 0
    \end{array} \right) \,,\quad \quad
 E_{\Lambda}=\left( \begin{array}{cc}
	0 & 1 \\
	1 & 0
    \end{array} \right)\,,
\end{equation}
where we have also written the $\Lambda$--sector vacuum matrix, $E_{\Lambda}$.  
The true $\SU(4)_Q$ vacuum can be written as a linear combination of the EW-preserving and EW-breaking vacua, 
${E_Q=c_\theta E_Q^-+s_\theta E_Q^{\mathrm{B}}}$. We use the short-hand notations $s_x\equiv \sin x, c_x\equiv \cos x$, and $t_x\equiv \tan x$ throughout.
Either choice of $E_Q^{\pm}$ is equivalent~\cite{Galloway:2010bp}, and in this letter, we have chosen $E_Q^-$.  

The Goldstone excitations around the vacuum are then parameterized by  
\begin{equation}
    \begin{split}
    \Sigma_Q &= \exp\left[ 2\sqrt{2}\, i  \left(\frac{\Pi_Q}{f}-\frac{1}{3}\frac{\Theta}{f_\Theta}\mathbb{1}_4\right) \right] E_Q\,, \\
    \Sigma_\Lambda &= \exp\left[ 2\sqrt{2}\,i \left( \frac{\Pi_\Lambda}{\fL}+\frac{1}{6}\frac{\Theta}{f_\Theta} \mathbb{1}_2\right)\right] 
    E_\Lambda\,, \quad 
    \end{split}
\label{eq:NGBmatrix}
\end{equation}
with
\begin{equation}
    \Pi_Q=\sum_{i=1}^5 \Pi_Q^ iX^i_Q , \quad\quad  \Pi_\Lambda=\sum_{a=1}^2 \Pi_\Lambda^a X^a_\Lambda\,,
\end{equation}
where $X_{Q,\Lambda}$ are the $\theta$-dependent broken generators of $ \SU(4)_Q $ and $ \SU(2)_\Lambda $ and can be found explicitly in Refs.~\cite{Ryttov:2008xe,Galloway:2010bp}.
The $\Theta$ state is the only state connecting the two sectors in the effective Lagrangian. The NGB matrices in Eq.~(\ref{eq:NGBmatrix}) will be modified to Eq.~(\ref{eq:NGBmatrixMod}) when we add the PC fermion mass operators in Eq.~(\ref{eq: PC operator top yukawa}). In the composite Higgs range, $ 0<\theta < \pi/2 $, we identify $h\equiv\Pi_Q^4 \sim \ct (\bar{U}U+\bar{D}D) + \st \,{\rm Re} \, U^TCD $  as the composite Higgs and $\eta\equiv \Pi_Q^5 \sim \,{\rm Im} \, U^TCD  $, while the remaining three $ \Pi_L^{1,2,3} $ are exact Goldstones eaten by the massive $ W^\pm $ and $ Z $. Furthermore, we identify the DM candidate $\Phi\equiv\frac{1}{\sqrt{2}}(\Pi_\Lambda^1-i \Pi_{\Lambda}^2)\sim \Lambda^T C \Lambda$ and $\bar{\Phi}\equiv\frac{1}{\sqrt{2}}(\Pi_\Lambda^1+i \Pi_{\Lambda}^2)\sim \overline{\Lambda}^T C \overline{\Lambda} $ in the $\Lambda$--sector, while $ \Theta\sim i (\overline{U}\gamma^5 U+\overline{D}\gamma^5 D-(1/2)\overline{\Lambda}\gamma^5 \Lambda) $. Note that, following Ref.~\cite{Alanne:2018xli}, we have here used Dirac spinors to indicate the hyper-fermions. 

The four-fermion interactions generating PC operators can have various forms. In this work, we are interested specifically on the ones in Eq.~(\ref{eq: PC operator top yukawa}). For these operators, we have imposed that only combinations of hyper-fermions that are $ \mathrm{U}(1)_\Lambda $ neutral couple to the top fields, so that this symmetry remains preserved such that the DM candidate is stable. One possible extension of this model example is to have six new $ \chi $ Weyl fermions in the two-index anti-symmetric representation ($ \mathbf{A_2} $) of $ G_{\rm HC}=\SP(2N)_{\rm HC} $, as detailed in Ref.~\cite{Ferretti:2014qta}, while the four $ Q $ and two $ \Lambda $ Weyl fermions are still in the fundamental ($ \mathbf{F} $) and adjoint representation ($ \mathbf{G} $), respectively. Asymptotic freedom is guaranteed by $ N<3 $, when the $ b $ coefficient of the beta function is
positive~\cite{Ryttov:2008xe,Barnard:2013zea}, i.e. $ b=-10(N-3)/3>0 $. Therefore, we choose that $ G_{\rm HC}=\SP(4)_{\rm HC} $. In this scenario, the global symmetry in Eq.~(\ref{eq: coset symmetry}) is extended to \begin{equation}
    \begin{split}
\frac{\SU(4)_Q \otimes \SU(2)_\Lambda\otimes \SU(6)_\chi \otimes \UU(1)_{\Theta_1}\otimes \UU(1)_{\Theta_2} }{\SP(4)_Q\otimes \UU(1)_{\Lambda}\otimes \mathbb{Z}_2\otimes \SO(6)_\chi} \,.
    \end{split}
\label{eq:NGBmatrixMod}
\end{equation} Now, there are two anomaly-free U(1) corresponding to two pNGBs, $ \Theta_{1,2} $, and one anomalous U(1) corresponding to the state $ \Theta' $. 
The $Q$, $\Lambda$ and $\chi$ charges under these U(1) are defined by the anomaly cancellation
\begin{equation}
q_Q T(\mathbf{F}) + q_\Lambda T(\mathbf{G}) +q_\chi T(\mathbf{A_2})=0\Rightarrow q_\Lambda = - \frac{1}{3}q_Q-q_\chi
\end{equation} with $T(\mathcal{R})=1,\,2N_{\mathrm{HC}}+2,\,2N_{\mathrm{HC}}-2$ the index of representation $\mathcal{R}={\bf F,\,G,\,A_2}$, respectively. In our case, we have $n=3$ fermionic sectors, $ Q $ in {\bf F}, $\Lambda$ in {\bf G} and $\chi$ in $\bf A_2$ of $G_{\mathrm{HC}}=\mathrm{Sp}(4)_{\mathrm{HC}}$. Therefore, the anomalous charges are $q_{F,3}=1$, $q_{G,3}=3$, $q_{A_2,3}=3$ and the charges of the two anomaly-free are the orthogonal combinations $ q_{F,1}=-3 $, $ q_{G,1}=1 $, $ q_{A_2,1}=0 $ and $ q_{F,2}=0 $, $ q_{G,2}=1 $, $ q_{A_2,2}=-1 $. By following Appendix~B in Ref.~\cite{Alanne:2018xli}, these modifications are incorporated by replacing the pNGB matrix in Eq.~(\ref{eq:NGBmatrix}) by \begin{equation}
    \begin{split}
    \Sigma_Q &= \exp\left[ 2\sqrt{2}\, i  \left(\frac{\Pi_Q}{f}-\frac{3}{2\sqrt{19}}\frac{\Theta_1}{f_{\Theta_1}}\mathbb{1}_4\right) \right] E_Q\,, \\
    \Sigma_\Lambda &= \exp\left[ 2\sqrt{2}\,i \left( \frac{\Pi_\Lambda}{\fL}+\frac{1}{2\sqrt{19}}\frac{\Theta_1}{f_{\Theta_1}} \mathbb{1}_2+\frac{1}{2}\frac{\Theta_2}{f_{\Theta_2}} \mathbb{1}_2\right)\right] 
    E_\Lambda\,, 
    \end{split}
\label{eq:NGBmatrixMod}
\end{equation} where the $ \Theta_2 $ state in the $ \chi $--sector is typically decoupled by adding an explicit mass, $ m_\chi $. 

\section{Effective Lagrangian and vacuum alignment}

Upon the compositeness scale, $ \Lambda_{\rm HC}\sim 4\pi f $, the effective Lagrangian can be written as
\begin{equation}
    \label{eq:effLag}
    \mathcal{L}_\mathrm{eff}=\mathcal{L}_{\mathrm{kin}}+\mathcal{L}_{\mathrm{f}}-V_{\mathrm{eff}}\,,
\end{equation}
where the kinetic terms are
\begin{equation}
    \label{eq:kinLag}
    \mathcal{L}_{\mathrm{kin}}=\frac{f^2}{8}\Tr [D_{\mu}\Sigma_Q^{\dagger}D^{\mu}\Sigma_Q]+\frac{\fL^2}{8}\Tr [\partial_{\mu}\Sigma_\Lambda^{\dagger}\partial^{\mu}\Sigma_\Lambda]
\end{equation}
with 
\begin{equation}
    \label{eq:covD}
    D_{\mu}\Sigma_Q=\partial_{\mu}\Sigma_Q-\ii\left(G_{\mu}\Sigma_Q+\Sigma_QG_{\mu}^{T}\right)\,,
\end{equation}
and the EW gauge fields are encoded in the covariant derivative
\begin{equation}
    \label{eq:Gmu}
    G_{\mu}=g_L W_{\mu}^iT_{\mathrm{L}}^i+g_Y B_{\mu}T_{\mathrm{R}}^3\,.
\end{equation} Besides providing kinetic terms and self-interactions for the pNGBs, it will induce masses for the EW gauge bosons and their couplings with the pNGBs (including the SM Higgs identified as $ h $), 
\beq \label{WZ masses and SM VEV}
&&m_W^2=\frac{1}{4}g_L^2f^2s_\theta^2\,,\quad\quad m_Z^2=m_W^2/c^2_{\theta_W}\,, \\
&&g_{hWW}=\frac{1}{2}g^2_Wfs_\theta c_\theta=g_{  hWW}^{\rm SM}c_\theta\,,\quad\quad g_{ hZZ}=g_{ hWW}/c^2_{\theta_W}\,, \nonumber
\eeq 
where $ v_{\rm SM}\equiv fs_\theta = 246~\text{GeV} $, $ g_{L,Y} $ are the electroweak $ \SU(2)_{\rm L} $ and $ \UU(1)_{\rm Y} $ gauge couplings, and $ \theta_W $ is the Weinberg angle. 
The vacuum misalignment angle $ \theta $ parametrizes the 
corrections to the Higgs couplings to the EW gauge bosons and 
is constrained by LHC data~\cite{deBlas:2018tjm}. 
This would require a small $ \theta $ ($ s_\theta \lesssim 0.3 $), 
however an even smaller value is needed by the EW precision 
measurements ($s_\theta \lesssim 0.2$). 

The PC four-fermion operators in Eq.~(\ref{eq: PC operator top yukawa}) contribute to $ \mathcal{L}_{\mathrm{f}} $ in Eq.~(\ref{eq:effLag}) with effective operators generating the top mass and Yukawa coupling, which can be written as~\cite{Alanne:2018wtp}
\begin{equation}
    \begin{split} \label{eq: top Yukawa operator}
     \mathcal{L}_{\mathrm{f}}&\supset \frac{C_{yS}}{4\pi} y_{L} y_{R}f (t_L t^c_R)^\dagger~{\rm Tr} [P_Q^1 \Sigma^\dagger_Q] {\rm Tr} [P_\Lambda\Sigma_\Lambda^\dagger] +\mathrm{h.c}\\
    &=(t_L t^c_R)^\dagger \left( m_\mathrm{top} + \frac{m_\mathrm{top}}{v_{\rm SM}} c_\theta h 
	 + \dots \right) +\mathrm{h.c.}\,, 
    \end{split}
\end{equation}
where $C_{yS}\sim\mathcal{O}(1)$,  $m_{\rm top} =C_{yS} y_{L} y_{R} v_{\rm SM} /(2\pi)$, and $ y_{L/R} $ are related to the couplings $ \widetilde{y}_{L/R} $ via the anomalous dimensions of the fermionic operators and are expected to be $ \mathcal{O}(1) $ for the top.

The value of $\theta$, and the amount of misalignment, is controlled by 
the effective potential  $V_{\mathrm{eff}}$ in Eq.~(\ref{eq:effLag}), which receives contributions from
 the EW gauge interactions, the vector-like masses of the hyper-fermions and the SM fermion couplings 
to the strong sector. At leading order, each source of symmetry breaking contributes independently to
the effective potential:
\beq
V_{\text{eff}}&=&V_{\text{gauge}}+V_\text{m} +V_{\text{top}}+V_\Lambda+\dotsc \,, \label{Potential 1}
\eeq 
where the dots are left to indicate the presence of mixed terms at higher orders,
or the effect of additional UV operators.
In this work, we will write the effective potential in terms of effective operators,
which contain insertions of spurions that correspond to the symmetry breaking 
couplings. A complete classification of such operators, for this kind of cosets, up
to next-to-leading order can be found in Ref.~\cite{Alanne:2018wtp}.

Both the contribution of gauge interactions and of the hyper-fermion masses arise
to leading $\mathcal{O} (p^2)$ order and have a standard form:
\begin{equation}
\begin{aligned} V_{\rm gauge,p^2}&=- C_g f^4 \left( \sum_{i=1}^3 g_L^2 \text{Tr}[T_L^i \Sigma_Q T_L^{iT}\Sigma_Q^\dagger]+g_Y^{2}\text{Tr}[T_R^3 \Sigma_Q T_R^{3T} \Sigma_Q^\dagger]\right) \,, \end{aligned} \end{equation} 
where the non-perturbative $ \mathcal{O}(1) $ coefficient $ C_g $ can be determined by the lattice simulations~\cite{Arthur:2016dir} and $T_{L/R}$ are given in Eq.~(\ref{eq:gensLR}), while \begin{equation}
\begin{aligned}
    \label{eq:V0}
   V_{\rm m,p^2}&=2\pi\cQ f^3\,\Tr\left[M_Q\Sigma_Q^{\dagger}+\Sigma_QM_Q^{\dagger}\right]
    +2\pi\cL \fL^3\,\Tr [M_{\Lambda}\Sigma_{\Lambda}^{\dagger}+\Sigma_{\Lambda}M_{\Lambda}^{\dagger}]\,,
\end{aligned}\end{equation} where $ M_Q={\rm diag}(m_1 i\sigma_2,-m_2 i\sigma_2) $ and  $ M_\Lambda=m \sigma_1 $ are the mass matrices of the $ \mathcal{R}_{1,2} $ hyper-fermions, respectively. The coefficients $\cQ$,$\cL$ are non-perturbative $\mathcal{O}(1)$ constants, and we use the numerical value ${\cQ\approx 1.5}$ suggested by the lattice simulations~\cite{Arthur:2016dir}. The mass terms involving $M_Q$ (as well as the subleading EW gauge interactions) prefer the vacuum where the EW is unbroken. The correct vacuum alignment must, therefore, be ensured by the SM-fermion mass generation mechanism. The top loop contributions arising from the four-fermion PC operator in Eq.~(\ref{eq: PC operator top yukawa}) yield at leading order (LO) in the chiral expansion the effective potential contribution as follows~\cite{Alanne:2018wtp}  \begin{equation}
\begin{split} \label{eq: LO terms}
V_{\rm top,p^2}&=\frac{C_L f^4}{4\pi} y_L^2  {\rm Tr}[P_Q^\alpha \Sigma_Q^\dagger]{\rm Tr}[\Sigma_Q P_{Q \alpha}^{\dagger}]+\frac{C_R f^4_\Lambda}{4\pi} y_R^2 {\rm Tr}[P_\Lambda \Sigma_\Lambda^\dagger]{\rm Tr}[\Sigma_\Lambda P_\Lambda^\dagger]\,.
    \end{split}
\end{equation} At next-to-leading order (NLO), these PC operators yield~\cite{Alanne:2018wtp}\allowdisplaybreaks 
\begin{equation} \begin{aligned}  \label{eq: NLO terms}
V_{\rm top,p^4}= & \frac{C_{LL}f^4}{(4\pi)^2}  y_L^4\bigg( {\rm Tr} [P_Q^\alpha \Sigma_Q^\dagger]^2 {\rm Tr} [\Sigma_Q P_{Q\alpha}^\dagger ]^2+{\rm Tr} [P_Q^\alpha \Sigma_Q^\dagger P_Q^\beta \Sigma_Q^\dagger]{\rm Tr} [\Sigma_Q P_{Q\alpha}^\dagger \Sigma_Q  P_{Q\beta}^\dagger ]+\\ &\phantom{++++++}\Big\lbrace{\rm Tr} [P_Q^\alpha \Sigma_Q^\dagger]{\rm Tr} [P_Q^\beta \Sigma_Q^\dagger]{\rm Tr} [\Sigma_Q P_{Q\alpha}^\dagger \Sigma_Q P_{Q\beta}^\dagger] +{\rm Tr} [P_Q^\alpha \Sigma_Q^\dagger] {\rm Tr} [P_Q^\beta P_{Q\alpha}^\dagger  \Sigma_Q P_{Q\beta}^\dagger] +{\rm h.c.}\Big\rbrace \bigg)+
\\ &\frac{C_{RR}f^4_\Lambda}{(4\pi)^2}  y_R^4\bigg( {\rm Tr} [P_\Lambda \Sigma_\Lambda^\dagger]^2 {\rm Tr} [\Sigma_\Lambda P_\Lambda^\dagger ]^2+{\rm Tr} [P_\Lambda \Sigma_\Lambda^\dagger P_\Lambda \Sigma_\Lambda^\dagger]{\rm Tr} [\Sigma_\Lambda P_\Lambda^\dagger \Sigma_\Lambda  P_\Lambda^\dagger ]+\\ &\phantom{++++++}\Big\lbrace{\rm Tr} [P_\Lambda \Sigma_\Lambda^\dagger]{\rm Tr} [P_\Lambda \Sigma_\Lambda^\dagger]{\rm Tr} [\Sigma_\Lambda P_\Lambda^\dagger \Sigma_\Lambda P_\Lambda^\dagger] +{\rm Tr} [P_\Lambda \Sigma_\Lambda^\dagger] {\rm Tr} [P_\Lambda P_\Lambda^\dagger  \Sigma_\Lambda P_\Lambda^\dagger] +{\rm h.c.}\Big\rbrace \bigg)+ \\  &\frac{C_{LR}f^2f_\Lambda^2}{(4\pi)^2} y_L^2 y_R^2{\rm Tr} [P_Q^\alpha \Sigma_Q^\dagger ] {\rm Tr} [ \Sigma_Q P_{Q\alpha}^\dagger] {\rm Tr} [P_\Lambda \Sigma_\Lambda^\dagger] {\rm Tr} [ \Sigma_\Lambda P_\Lambda^\dagger]  \,, \end{aligned} \end{equation}  where $ C_{L,R} $ and $ C_{LL,RR,LR} $ are non-perturbative $ \mathcal{O}(1) $ constants. Now, we have introduced the LO and NLO potential terms to the top potential $ V_{\rm top}=V_{\rm top,p^2}+V_{\rm top,p^4}+\dots $, which contributes to the total effective Lagrangian $ V_{\rm eff} $ in Eq.~(\ref{eq:effLag}). Finally, it is relevant for the DM mass to consider possible four-hyper-fermion operators in the $\Lambda$--sector of the form $ \Lambda \Lambda \Lambda \Lambda $, which yield potential contributions~\cite{Foadi:2007ue}:
\begin{align}
V_{\Lambda}&=C_\Lambda g_\Lambda^2 f^4_\Lambda  \Tr[\sigma_3\Sigma_\Lambda^{\dagger}\sigma_3\Sigma_\Lambda]\,,
\label{eq: Lambda4 term}
\end{align} which preserves the symmetry $ \mathrm{U}(1)_\Lambda $ and has no effects on the vacuum alignment. The coefficient $ C_\Lambda $ is $ \mathcal{O}(1) $ form factor that can be computed on the lattice~\cite{Arthur:2016ozw}.

To zeroth order depending on vacuum alignment angle $ \theta $, the total potential in Eq.~(\ref{Potential 1}) is given by \begin{equation}
\begin{split}
V_{\rm eff}^0(\theta)=&- C_g f^4 \frac{3g_L^2+g^{2}_Y}{2}c_\theta^2+8\pi f^3  c_Q  m_Q c_\theta +\frac{f^4}{2\pi}C_L y_L^2s_\theta^2  \\& + \frac{f^4}{(4\pi)^2}\Big[\left(3C_{LL}y_L^4+4C_{LR}y_L^2y_R^2 f_\Lambda^2/f^2\right)s_\theta^2+9C_{LL}y_L^4 s_\theta^4\Big] +\dots\,, \end{split}
\end{equation} where $ m_Q\equiv m_1+m_2 $. By minimizing this potential, $ \partial V_{\rm eff}^0/\partial \theta = 0 $, we can fix the hyper-fermion mass term as a function of the misalignment angle (or vice versa) as follows\begin{equation}
    \label{eq:alignPC}
    m_Q=\frac{f c_\theta}{64\pi^3 c_Q }\bigg(3C_{LL}y_L^4(4-3 c_{2\theta})+4C_{LR}y_L^2 y_R^2 f_\Lambda^2/f^2+8\pi C_L y_L^2 +8\pi^2 C_g(3g_L^2+g_Y^2)\bigg)\,.
\end{equation} 

\section{The pNGB mass expressions}

From the total effective potential in Eq.~(\ref{Potential 1}) and the above vacuum misalignment condition, the physical SM Higgs, $ h $, obtains a mass
\begin{align}
    \label{eq: higgs mass}
    m_h^2=\frac{v_{\rm EW}^2}{8\pi^2}\bigg( 3 C_{LL} y_L^4 (2+9 c_{2\theta})- 4 C_{LR} y_L^2 y_R^2 f_\Lambda^2/f^2 - 8 \pi C_L y_L^2 -8\pi^2 (3g_L^2+g_Y^2) \bigg)\,.
\end{align} Considering the following form of the above Higgs mass expression up to NNNLO: \begin{align}
  m_h^2 \approx ( -0.32 C_L y_L^2 + 0.42 C_{LL} y_L^4+ 0.097 C_{LLL} y_L^6 s_\theta^2 + 0.031 C_{LLLL} y_L^8 s_\theta^4 + \dots )v_{\rm SM}^2
\end{align} with $ y_L\sim 2.1 $ fixed by the top mass in Eq.~(\ref{eq: top Yukawa operator}), if we assume $ y_R=y_L $ and $ C_{yS}\sim 1 $. The dots include the $ C_{LR} $ and $ C_g $ terms along with the $ C_L $ higher order terms, sub-dominating relative to the $ C_{LL} $ terms. 
For $ y_L\gtrsim 0.9 $, the squared Higgs mass is assured to be positive due to the fact that the NLO potential terms dominate over the LO terms. Furthermore, the higher order potential terms, with the coefficients $  C_{LLL}$, $ C_{LLLL},\dots $, are suppressed by a small $ s_\theta \ll 1/y_L \sim 0.5 $ relative to both the LO and NLO terms. Therefore, we can obtain $ m_h^2 >0 $ for $ y_L=y_R\sim 2.1 $ and $ C_i =\mathcal{O}(1) $.

Moreover, the mass of the DM candidate, identified with the pNGB $ \Phi $, is given by \begin{equation}
    \label{eq:}
    m_{DM}^2= \frac{1}{2\pi^2}\bigg(32\pi^2 C_\Lambda g_\Lambda^2 f_\Lambda^2+32\pi^3 c_\Lambda f_\Lambda m-13 C_{RR} y_R^4 f_\Lambda^2 - 2C_{LR} y_L^2 y_R^2 f^2 s_\theta^2-4\pi C_R y_R^2 f_\Lambda^2 \bigg)\,.
\end{equation} 

Finally, the last three pNGBs, $ \eta $ and $ \Theta_{1,2} $, in Eq.~(\ref{eq:NGBmatrixMod}) mix with each other, resulting in the squared mass matrix in the basis $ (\eta,\Theta_1,\Theta_2) $:  \begin{equation}
    \begin{split} \label{eq: mass matrix Theta eta}
&M_{\eta\Theta}^2=
\begin{pmatrix}
m_\eta^2 & m_{\eta\Theta_1}^2 & 0 \\
m_{\Theta_1\eta}^2 & m_{\Theta_1}^2  & m_{\Theta_1\Theta_2}^2
\\
0 & m_{\Theta_2\Theta_1}^2  & m_{\Theta_2}^2
\end{pmatrix}
    \end{split}
\end{equation} with \begin{equation*}
 \begin{split}
&m_\eta^2=8 \pi c_Q f m_Q/c_\theta\,, \quad\quad m_{\Theta_1}^2=\frac{16}{19}\frac{1}{f_{\Theta_1}^2}\pi (9 f^3 c_Q m_Q c_\theta+f_{\Lambda}^3 c_\Lambda m)\,, \quad\quad m_{\Theta_1\eta}^2 = m_{\eta\Theta_1}^2  =-24\sqrt{\frac{2}{19}} \frac{f^2}{f_{\Theta_1}} \pi c_Q  (m_1-m_2)\,, \\ &  m_{\Theta_2}^2=16\frac{f_{\Lambda}^3}{f_{\Theta_2}^2}\pi  c_\Lambda m +m_\chi^2\,, \quad \quad m_{\Theta_1\Theta_2}^2=m_{\Theta_2\Theta_1}^2 = \frac{16}{\sqrt{19}} \frac{f_{\Lambda}^3}{f_{\Theta_1}f_{\Theta_2}}\pi  c_\Lambda m \,.
    \end{split}
\end{equation*} If we assume that the mass $ m $ of the hyper-fermions in the $ \Lambda $--sector is vanishing, the pNGB $ \Theta_2 $ is decoupled from the other two pNGBs and its mass is the explicit mass $ m_\chi $. Moreover, if one of the vector-like masses, $ m_1 $ or $ m_2 $, of the hyper-fermions in the $ Q $--sector is vanishing, the mass eigenstate $ \Theta $, consisting mostly of $ \Theta_1 $, is massless although $ f_{\Lambda}\neq f_{\Theta_1}\neq f $, while $ \widetilde{\eta} $, consisting mostly of $ \eta $, has a mass of order $ f $ for $ f_{\Theta_1}\sim f $: \begin{equation}
    \begin{split}
m_{\Theta}^2=0 \,,\quad \quad \quad m_{\widetilde{\eta}}^2 = \frac{8}{19}\frac{18f^2 c_\theta^2+19f_{\Theta_1}^2}{f_{\Theta_1}^2c_\theta}\pi c_Q f m_Q \,. 
    \end{split}
\end{equation} 

The kinetic term of $\Theta_{1,2}$ in Eq.~(\ref{eq:kinLag}) are canonically normalized only if $f_{\Lambda}=f_{\Theta_1}=f_{\Theta_2}\equiv f$, which we will assume for simplicity. In the general case, the kinetic terms must be renormalized, but based on Casimir scaling we expect them to be of the same size~\cite{Ryttov:2008xe,Frandsen:2011kt}. Assuming either $ m_1 $ or $ m_2 $ is vanishing, $ m=0 $ and $f_{\Lambda}=f_{\Theta_1}=f_{\Theta_2}\equiv f$, the mass expressions of the pNGBs in this model example are given by \begin{align} \label{eq: mass expressions of pNGBs}
    &m_h^2=\frac{v_{\rm EW}^2}{8\pi^2}\bigg( 3 C_{LL} y_L^4 (2+9 c_{2\theta})- 4 C_{LR} y_L^2 y_R^2 - 8 \pi C_L y_L^2 -8\pi^2 (3g_L^2+g_Y^2) \bigg)\,, \nonumber \\ &m_{DM}^2= \frac{f^2}{2\pi^2}\bigg(32\pi^2 C_\Lambda g_\Lambda^2-13 C_{RR} y_R^4  - 2C_{LR} y_L^2 y_R^2  s_\theta^2-4\pi C_R y_R^2 \bigg)\,, \\
   & m_{\Theta} \simeq 0,\quad \quad \quad m_{\Theta_2}=m_\chi \,, \quad \quad \quad m_{\widetilde{\eta}}^2 = \frac{8}{19}(28+9c_{2\theta}) \pi c_Q f m_Q/c_\theta = \frac{296}{19}\pi c_Q f m_Q +\mathcal{O}(s_\theta^2)\,. \nonumber
\end{align}  However, the $ \Theta $ can achieve a small mass from its mixing with the $ \Theta' $ state corresponding to the $ \mathrm{U}(1) $ symmetry which is anomalous. In addition, the mass of $ \Theta' $ is generated by instanton effects related to the $ \mathrm{U}(1) $ anomaly~\cite{Belyaev:2016ftv}.  

\section{The self-interacting Dark matter candidate}

To obtain strong enough elastic scattering among DM themselves, the DM candidate requires a mass of $ m_{DM}\lesssim \mathcal{O}(1)~\text{GeV}$, which are very small relative to the Higgs decay constant of $ f \gtrsim \mathcal{O}(1)~\text{TeV}$. Since $ m_{DM}^2 $ is negative when $ C_B g_B^2 =0 $, a small DM mass, $ m_{DM} \ll f $, can be achieved by tuning $ C_B g_B^2 $ to a certain value of order unity. In this scenario, the efficient annihilation mode of the DM candidate involves a pair of pNGBs that do not carry $ \mathrm{U}(1)_\Lambda $-charges: by studying the potential, we found that the dominant channel involves the $\mathrm{U}(1)_\Theta$ pNGB $\Theta$, which can be parametrically lighter than the other pNGBs, see Eq.~(\ref{eq: mass expressions of pNGBs}). Thus, the dominant annihilation channel is the contact interaction $ \Phi + \overline{\Phi} \rightarrow \Theta +\Theta$ with the coupling $ \lambda_{\Phi \Phi \Theta \Theta} $ given by \begin{equation}
    \begin{split}
\mathcal{L}_{\rm eff}\supset  \lambda_{\Phi \Phi\Theta\Theta}\Phi \overline{\Phi} \Theta \Theta 
    \end{split}
\end{equation} with \begin{equation}
    \begin{split}
\lambda_{\Phi\Phi\Theta \Theta }=\frac{18 C_{LR} y_L^2 y_R^2 s_\theta^2 c_\theta^2}{(19+18 c_\theta^2)\pi^2}= \frac{18C_{LR} y_L^2 y_R^2}{37\pi^2}s_\theta^2+\mathcal{O}(s_\theta^4)\,,
    \end{split}
\end{equation}  where $ s_\theta \ll 1 $. This quartic coupling is suppressed by the misalignment in the Q--sector, $ \lambda_{\Phi \Phi\Theta\Theta}\approx \lambda_0 (v_{\rm EW}/f)^2 =\lambda_0 s_\theta^2  $. The expression of $ \lambda_0 $, mentioned in the main text, is, therefore, given by \begin{equation} \begin{aligned}\label{eq: lambda0}
\lambda_0\approx\frac{18C_{LR} y_L^2 y_R^2 }{37\pi^2} \,,
\end{aligned}\end{equation} which can be of order unity. 

Furthermore, the coupling of the DM self-interactions is given by\begin{equation}
    \begin{split}
\mathcal{L}_{\rm eff}\supset  \lambda_{\Phi}(\Phi \overline{\Phi})^2
    \end{split}
\end{equation} with \begin{equation}
    \begin{split}
\lambda_{\Phi }=\frac{16C_{RR}y_R^4}{\pi^2}-\frac{4 m_{DM}^2}{3f^2}\approx \frac{16C_{RR}y_R^4}{\pi^2} 
    \end{split}
\end{equation} for $ m_{DM}\ll f $.

Finally, we consider a viable benchmark point in the parameter space of this model example: Assuming $ y_L=y_R \equiv y_t $ in Eq.~(\ref{eq: top Yukawa operator}), we obtain $ y_t\approx 2.10/\sqrt{C_{yS}} $ from the expression of the top mass in Eq.~(\ref{eq: top Yukawa operator}). Thus, the intervals of $ f $ and $ m_{DM} $  \begin{equation} \begin{aligned}
&16.5~\text{TeV} \lesssim f \lesssim 24.2~\text{TeV}, \quad\quad 0.20~\text{GeV} \lesssim m_{DM} \lesssim 0.42~\text{GeV}\,,
\end{aligned}\end{equation} give rise to a DM relic density $ \langle  \sigma v\rangle_{\Theta} \approx 5\times 10^{-26}~\text{cm}^3\text{s}^{-1}$~\cite{Steigman:2012nb} and strong self-interactions among the DM particles with cross-section per mass in the range $ \sigma_{SI}/m_{DM} \sim 0.1-1~\text{cm}^2/\text{g} $~\cite{Tulin:2017ara}. To obtain the observed Higgs mass and DM masses in the above intervals, when $ C_{LR}=C_{RR}=C_L=C_R=C_\Lambda=C_{yS}=1 $, we need $ C_{LL}\approx 0.50 $ and $ g_\Lambda \approx 0.99 $, respectively, given by the mass expressions in Eq.~(\ref{eq: mass expressions of pNGBs}).

\end{document}